\documentclass[aps,prx,twocolumn,groupedaddress]{revtex4-2}

\usepackage[utf8]{inputenc}
\usepackage{amssymb}
\usepackage{amsmath}
\usepackage{amsfonts}
\usepackage{mathrsfs}
\usepackage{color}
\usepackage{dsfont}
\usepackage{graphicx}
\usepackage{dcolumn}
\usepackage{bm}
\usepackage{hyperref}

\begin{document}
\title{Dynamical mean-field theory for the Hubbard-Holstein model on a quantum device}

\author{Steffen Backes$^{1,2,3}$}
\email{steffen-backes@g.ecc.u-tokyo.ac.jp}
\author{Yuta Murakami$^{2}$}
\author{Shiro Sakai$^{2}$}
\author{Ryotaro Arita$^{1,2}$}
\affiliation{$^1$Research Center for Advanced Science and Technology, University of Tokyo, Komaba, Tokyo 153-8904, Japan}
\affiliation{$^2$Center for Emergent Matter Science, RIKEN, Wako, Saitama 351-0198, Japan}
\affiliation{$^3$CPHT, CNRS, École polytechnique, Institut Polytechnique de Paris, 91120 Palaiseau, France}
\date{\today}

\begin{abstract}
Recent developments in quantum hardware and quantum algorithms have made it possible to utilize the capabilities of current noisy intermediate-scale quantum devices for addressing problems in quantum chemistry and condensed matter physics. Here we report a demonstration of solving the dynamical mean-field theory (DMFT) impurity problem for the Hubbard-Holstein model on the IBM 27-qubit Quantum Falcon Processor \textit{Kawasaki}, including self-consistency of the DMFT equations. This opens up the possibility to investigate strongly correlated electron systems coupled to bosonic degrees of freedom and impurity problems with frequency-dependent interactions. The problem involves both fermionic and bosonic degrees of freedom to be encoded on the quantum device, which we solve using a recently proposed Krylov variational quantum algorithm to obtain the impurity Green's function. We find the resulting spectral function to be in good agreement with the exact result, exhibiting both correlation and plasmonic satellites and significantly surpassing the accuracy of standard Trotter-expansion approaches. Our results provide an essential building block to study electronic correlations and plasmonic excitations on future quantum computers with modern {\it ab initio} techniques.
\end{abstract}

\maketitle

% Main Text %%%%%%%%%%%%%%%%%%%%%%%%%%%%%%
\section{\label{sec:introduction}Introduction}
Recent developments on quantum and quantum-classical hybrid algorithms as well as in the continuous increase in computational capabilities have enabled significant progress in simulating interacting fermionic systems on quantum computers\cite{Abrams1997,Ortiz2001,Somma2003,Whitfield2011,Peruzzo2014,McClean2016,Arute2020,Huggins2022}.
Such simulations have a fundamental importance in condensed matter physics, quantum chemistry, and material science, since the exponentially increasing dimension of the Hilbert space has limited classical computations to rather small systems unless crude approximations are made. The algorithms developed for these classical computations, however, may benefit from an implementation on quantum computers. An example of such algorithms is the dynamical mean-field theory (DMFT)\cite{Metzner1989,Georges1992,GeorgesDMFT1996,Vollhardt2012}. 
Feasibility of employing quantum computers for DMFT  has already been demonstrated by simulations and actual implementations on noisy intermediate-scale quantum (NISQ) devices\cite{Bauer2016,Kreula2016,Rungger_dmft_qc, Keen_dmft_qc,Jaderberg_2020}. 
However, the accuracy of an implementation on quantum devices is limited by the current NISQ devices, which allow only for shallow circuits and exhibit a significant level of noise due to possible bit-flips or phase-flips\cite{Shor1995,Fowler2012,Preskill2018quantumcomputingin,Huang2019,Chen2021}. This places a significant constraint on methods such as the Trotter expansion, which often requires deep circuits with several controlled gate operations\cite{Bauer2016,Kreula2016,Chiesa2019}.
To reduce circuit complexity compared to Trotter expansion approaches and improve the accuracy of the Green's function obtained on quantum devices, different algorithms have been proposed. Real-time approaches such as the variational quantum simulation\cite{Li2017,Yuan2019,McArdle2019,Heya2019,Endo2020} or variational Hamiltonian ansatz (VHA)\cite{Wecker2015,Reiner_2019,Libbi2022} are based on obtaining a variational form of the time evolution operator, while the Lehmann-representation\cite{Rungger_dmft_qc,Endo2020}, quantum embedding\cite{Lupo2021,Vorwerk2022}, or Krylov techniques\cite{Jamet2021,Jamet2022} obtain the Green's function directly in the frequency representation.  The recently proposed Krylov variational quantum algorithm (KVQA)\cite{Jamet2021,Jamet2022} has been shown in simulations to be a promising candidate for solving the impurity problem related to DMFT on quantum hardware, as it requires significantly shallower circuits than other approaches.

At the same time, the DMFT framework has been extended to include nonlocal interactions and screening effects\cite{Kotliar2002,Biermann2003,Biermann2014,Nilsson2017,Lichtenstein2012} in terms of a frequency dependent effective interaction\cite{Aryasetiawan2004}. This model with a dynamical interaction can be represented in terms of a Hubbard-Holstein impurity model, where the impurity electrons are coupled to bosonic degrees of freedom. These extensions of DMFT pose an even greater computational challenge, and thus can be promising candidates to benefit from utilizing quantum algorithms implemented on future quantum computers. Although the implementation of these methods is a necessary step to reach future applications to more realistic Hamiltonians, it has not yet been done so far.

Here we present, to the best of our knowledge, 1) the first implementation of the KVQA on a current NISQ device, the IBM 27-qubit Quantum Falcon Processor \textit{Kawasaki}, and 2) the first implementation of the DMFT impurity problem for the Hubbard-Holstein model on a quantum computer to obtain the impurity Green's function for an electronic system coupled to bosonic degrees of freedom. Our investigation paves the way for possible applications of \textit{ab initio} computational methods for strongly correlated electron systems on future quantum computing devices.

The manuscript is structured as follows: we first present the model subject to our study and the formalism to obtain the impurity Green's function on the quantum device. Next, we present our results obtained on the \textit{Kawasaki} quantum processor and compare to the exact solution. The last section of this manuscript concludes with summary of the main points of our work.

%%%%%%%%%%%%%%%%%%%%%%%%%%%%%%
\section{\label{sec:model_formalism} Model \& Formalism}
The DMFT approximation is based on representing the local Green's function of a lattice system of interacting electrons by a single-site impurity model, coupled to noninteracting bath degrees of freedom. The main challenge lies in solving the interacting impurity problem, which despite the reduction to a local model remains a formidable many-body problem.
Here we study a two-site DMFT impurity problem of the Hubbard-Holstein model at half-filling, consisting of one interacting  impurity site coupled to a bosonic degree of freedom, and one non-interacting bath-site at zero energy which can exchange electrons with the impurity site. The system is given by the Hamiltonian
\begin{align}
    H
    =& Un_{\uparrow}n_{\downarrow} 
    + V\sum_{\sigma} ( c^{\dagger}_{\sigma}d_{\sigma} + d^{\dagger}_{\sigma}c_{\sigma} )
    -\mu ( n_{\uparrow} + n_{\downarrow}  ) \nonumber \\
    &+ \omega_0 b^{\dagger} b 
    + \lambda (b^{\dagger} + b)( n_{\uparrow} + n_{\downarrow}  ),
    \label{eq:Hubbard_Holstein_Hamiltonian}
\end{align}
where $c^{\dagger},c$ and $d^{\dagger},d$ correspond to the impurity and bath electronic creation/annihilation
operators, respectively, with the density $n_{\sigma} = c^{\dagger}_{\sigma}c_{\sigma}$ for the spin $\sigma$. $\mu$ is the impurity potential, $b^{\dagger},b$ represent the bosonic  creation/annihilation operators, 
$U$ is the local Coulomb interaction, $V$ the hybridization amplitude, $\omega_0$ the energy of the bosonic mode, and $\lambda$ the coupling strength to the fermionic degrees of freedom.

Using the Jordan-Wigner transformation\cite{JordanWigner,Whitfield2011}, we represent the electronic creation and annihilation operators by corresponding Pauli operators $X,Y,Z$ via
\begin{align}
    c^{\dagger}_i &= Z_0 \otimes ... Z_{i-1}\otimes (X_i - iY_i)/2, \\
    c_i &= Z_0 \otimes ... Z_{i-1}\otimes (X_i + iY_i)/2 ,
\end{align}
where the  index $i$ labels the different flavors (bath/impurity site and possible spin). Introducing the Pauli $Z$ operators acting on all flavors $j<i$ ensures the fermionic commutation relations $\{ c_i, c^{\dagger}_j \} = \delta_{ij}$. For the bosonic degrees of freedom, which in general involve an infinite number of  possible excitations, one has to introduce a cutoff in practice. This cutoff depends on the number of qubits available for encoding the bosonic states, and the level of noise of the quantum hardware, as an increasing number of bosonic excitations lead to more complex quantum circuits. Different types of bosonic encoding have been discussed, which aim at either reducing gate complexity or the number of qubits\cite{Somma2003,Veis2016,McArdle2019_vibrations}, and have been also applied to simulating systems with both fermionic and bosonic degrees of freedom such as the Holstein polaron problem\cite{Macridin2018A,Macridin2018}. Here we restrict our simulation to one possible boson, represented by one additional qubit, which is an approximation justified for large bosonic frequency $\omega_0$. In this case the different encodings become equivalent. This leads to the following representation for the bosonic operators
\begin{align}
b^{\dagger}b &= (I-Z)/2, \\
b^{\dagger} + b &= X ,
\end{align}
where $I$ is the identity operator.
Due to the bosonic commutation relations no additional padding with $Z$ operators is needed. 

With these transformations we map the Hamiltonian in Eq.\eqref{eq:Hubbard_Holstein_Hamiltonian} on five qubits $(q_1,q_2,q_3,q_4,q_5)$, using the ordering $(\uparrow_{imp}, \downarrow_{imp}, \uparrow_{bath} , \downarrow_{bath}, B )$, i.e. the first two qubits represent the impurity spin up/down component, the third and fourth qubits represent the bath spin up/down component, and the last qubit represents the boson. The resulting Hamiltonian takes the form
\begin{align}
H
=& (U/4 -\mu + \omega_0/2) IIIII 
+ \frac{U}{4} ZZIII \nonumber \\
& -\frac{V}{2}\left( XZXII + YZYII + IXZXI + IYZYI \right) \nonumber \\
& -\frac{\omega_0}{2}IIIIZ + \lambda IIIIX.
\label{eq:hamiltonian_with_pauli_operators}
\end{align}
Terms with vanishing expectation value at half filling such as $ZIIII$ have been dropped. To obtain the approximate ground state for this Hamiltonian we use the  variational quantum eigensolver (VQE) approach\cite{Peruzzo2014,Yung2014,McClean2016,Tilly2022} with a hardware-efficient ansatz\cite{Kandala2017} that exploits the symmetry properties of the wave function to reduce the number of gates as much as possible (see further explanation in the results section), as shown in Fig.\ref{fig:vqe_circuit}.

To obtain the impurity Green's function we make use of ibm\_kawasaki, an IBM Quantum System One with a 27-qubit Falcon R5.1 processor, and the recently proposed Krylov variational quantum algorithm (KVQA)\cite{Jamet2021,Jamet2022}. Compared to other approaches such as the Trotter decomposition or variational algorithms that obtain the Green's function on the time axis, or the Lehmann representation that requires the calculation of excited states, the KVQA has been shown in simulations to require significantly shallower circuits with less controlled gate operations. The procedure gives the retarded Green's function via a continuous fraction expansion, which for the half-filled particle-hole symmetric case can be written as\cite{GeorgesDMFT1996,Jamet2021}
\begin{align}
G(z) &= \frac{1}{2}\frac{1}{z - a_0 - \frac{b_1^2}{ z - a_1 - \frac{b_2^2}{...} } },
\end{align}
for each electron and hole part. The coefficients $a_n,b_n$ are generated by the Krylov algorithm as
\begin{align}
 b_n^2 &= \langle \chi_{n-1} | H^2 | \chi_{n-1} \rangle - a_n^2 - b_{n-1}^2 , \\
 | \chi_n \rangle &= \frac{1}{b_n} \left( H| \chi_{n-1} \rangle - a_{n-1}|\chi_{n-1} \rangle - b_{n-1}|\chi_{n-2} \rangle \right), \\
 a_n &= \langle \chi_n | H |  \chi_n \rangle,
\end{align}
where we use $|\chi_0 \rangle_- = c|\mathrm{GS} \rangle$ ($|\chi_0 \rangle_+ = c^{\dagger}|\mathrm{GS} \rangle$) for 
the occupied(unoccupied) part of the spectral function, given by $-\mathrm{Im}G(\omega  +i\delta +E_0)/\pi$, with $\delta > 0$ being a small convergence parameter. These expressions can be efficiently evaluated on a quantum computer if generating circuits for the Krylov basis states $|\chi_n \rangle$ can be found. Here we use the approach outlined in Ref.\cite{Jamet2021}, which is based on a variational procedure to find the optimal circuit that generates a quantum state maximizing the overlap with $|\chi_n \rangle$.

%%%%%%%%%%%%%%%%%%%%%%%%%%%%%%
\section{\label{sec:results}Results}
\begin{figure}[t]
    \centering
    \includegraphics[width=0.35\textwidth]{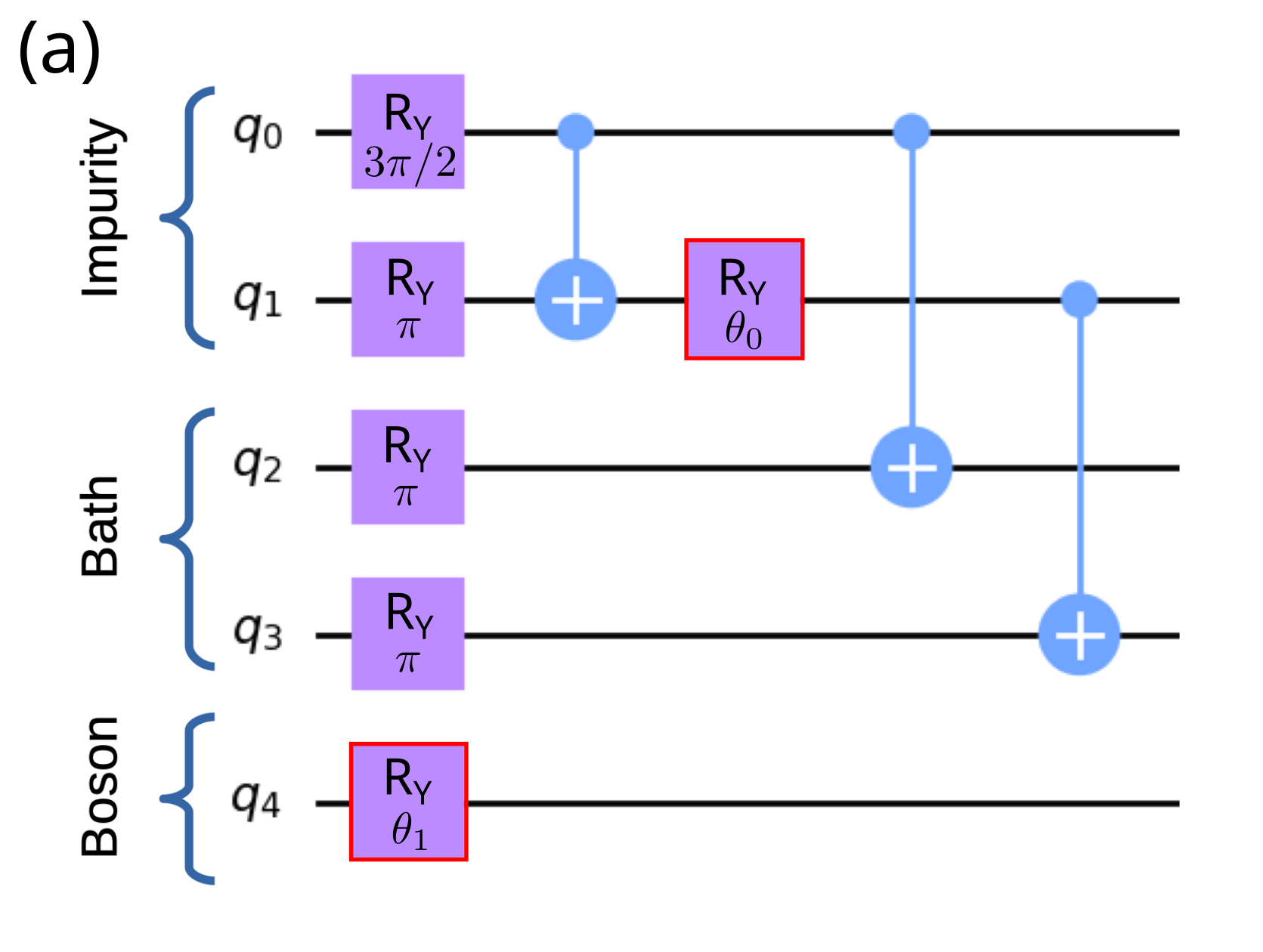}
    \includegraphics[width=0.5\textwidth]{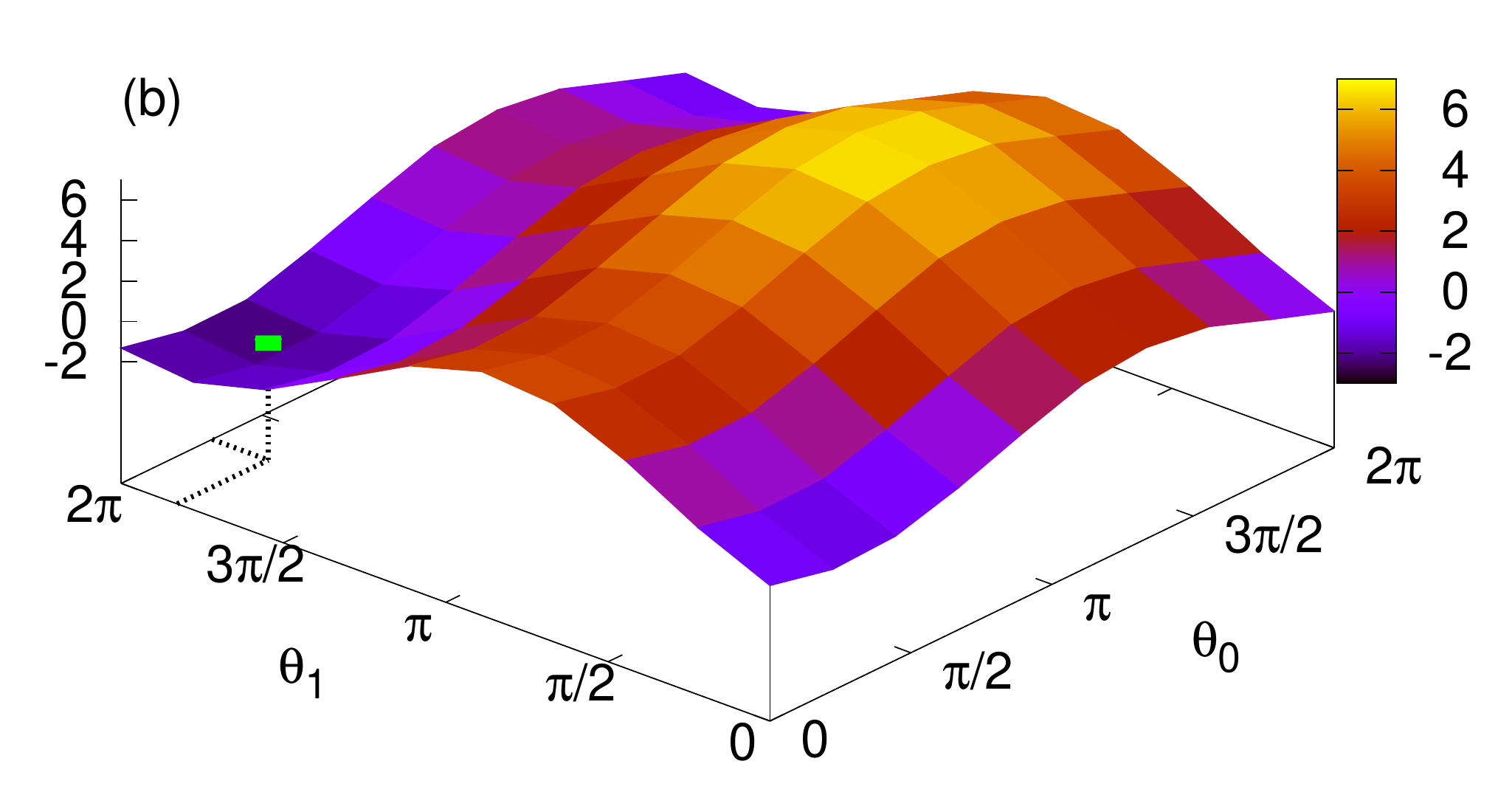}
    \caption{(a) The variational quantum eigensolver circuit for generating the ground state wave function
    of the Hubbard-Holstein DMFT impurity problem from Eq.\eqref{eq:hamiltonian_with_pauli_operators}. The ansatz wave function is parameterized by two rotation angles $\theta_0$   and $\theta_1$, indicated by the two red framed $R_Y$ rotation gates.
    (b) The energy landscape of the expectation value of the Hamiltonian operator measured on the \textit{Kawasaki}  quantum computer, depending on the  parameters $\theta_0$ and $\theta_1$. 
    The minimum around $(\theta_0,\theta_1)\approx (1.0, 5.7)$ (indicated by the green dot) corresponds to the approximate ground state energy $E_0= -2.43$, which is close to the exact value of $E_{0,exact}=-2.62$.
    }
    \label{fig:vqe_circuit}
\end{figure}

In the first step we will discuss the results for the impurity Green's function obtained for a fixed set of parameters, and in the second step we will demonstrate that the approach is robust enough to obtain a reliable DMFT  self-consistent solution on the Bethe lattice.

We first generate the ground state of the impurity problem using the VQE circuit as shown in  Fig.\ref{fig:vqe_circuit} (a) for the parameters $U=4, \omega_0=5, \lambda=1.5$ and $V=0.8$, which are representative for the self-consistent DMFT solution, as discussed below. By exploiting the symmetry properties of the impurity problem at half filling, we reduce the VQE circuit to a simplified ansatz for the ground state wave function of the following form
\begin{align}
    |\psi \rangle =& \left[ \sin\theta_0\left( |\uparrow\downarrow , 0\rangle+|0,\uparrow\downarrow \rangle \right)
                                + \cos\theta_0\left( |\uparrow, \downarrow\rangle -|\downarrow, \uparrow\rangle \right) 
 \right] \nonumber \\
                    &\otimes \left( \cos\theta_1 |0_B \rangle +  \sin\theta_1 |1_B \rangle \right).
\label{eq:ansatz}                    
\end{align}
This form is only approximately able to represent the true ground state wave function but allows for a parametrization of  the ansatz wave function and all resulting quantities such as the total energy in terms of only two parameters $\theta_0, \theta_1$. This enables us to resort to a two-dimensional scan of the energy landscape to reliably find the lowest energy and thus the best ground state approximation without relying on an optimization procedure that is hampered by local minima and barren plateaus\cite{McClean2018,Wang2021,Cerezo2021}. The resulting energy potential landscape obtained on the \textit{Kawasaki} quantum processor is shown in Fig.\ref{fig:vqe_circuit} (b). We find an energy minimum of $E_0= -2.43$ and thus the best possible approximation to the ground state around $(\theta_0,\theta_1)= (1.0, 5.7)$, which is close to the exact ground state energy $E_{0,exact}=-2.62$ for the system with the same bosonic cutoff.  The resulting ground state energy using the noise-free Qiskit simulator environment\cite{Qiskit}, obtained as $E_{0,sim}=-2.58$, is in close agreement with the exact value, indicating that the overestimation of the ground state energy on \textit{Kawasaki} is induced to a large degree by the noise of the device rather than the two-parameter approximation used in Eq.\eqref{eq:ansatz}.  Using error mitigation techniques\cite{Nation2021} should thus be a promising way for further improvements on the ground state energy.

We use the same approach of scanning the two-dimensional parameter space for obtaining the Krylov states and eventually the retarded Green's function and spectral function of the impurity site. First, we find the parameters for a circuit that approximates the state $| \chi_0\rangle = c_{\uparrow,imp}|\mathrm{GS}\rangle$, which is composed of the state $|\downarrow,0,0_B \rangle,|0, \downarrow,0_B \rangle, |\downarrow,0,1_B \rangle,|0, \downarrow,1_B \rangle$ by using a similar hardware efficient ansatz parametrized by two rotation angles. The final parameters are given by the ones that maximize the overlap with $| \chi_0\rangle$. The best candidate state that we find has an overlap of approximately $0.95$ (overlap of $1$ signals exact representation) with $| \chi_0\rangle$, but we point out that this error is composed both of the error in finding the right circuit parameters, and in the measurement of the overlap itself, i.e. even an exact representation will not result in an overlap of $1$ because of the measurement error on the quantum device.

Then we progressively measure the expectation values for $a_n, b^2_n$, and construct the circuit for next Krylov state $| \chi_n\rangle$, as outlined in Ref.\cite{Jamet2021}. Exploiting particle-hole symmetry at half-filling, we can obtain the unoccupied part of the spectrum from the same $a_n, b^2_n$ parameters obtained for the occupied part. We find that we can reliably obtain the Krylov states up to $n=2$ on the quantum computer. With further iterations the signal-to-noise ratio for $b_n^2$ becomes too small, often resulting in small negative values. Therefore, we obtain the Green's function from the parameters for the first two Krylov iterations, which generates up to three possible peaks in each the occupied and unoccupied part of the spectrum. 

\begin{figure}[t]
    \includegraphics[width=0.5\textwidth]{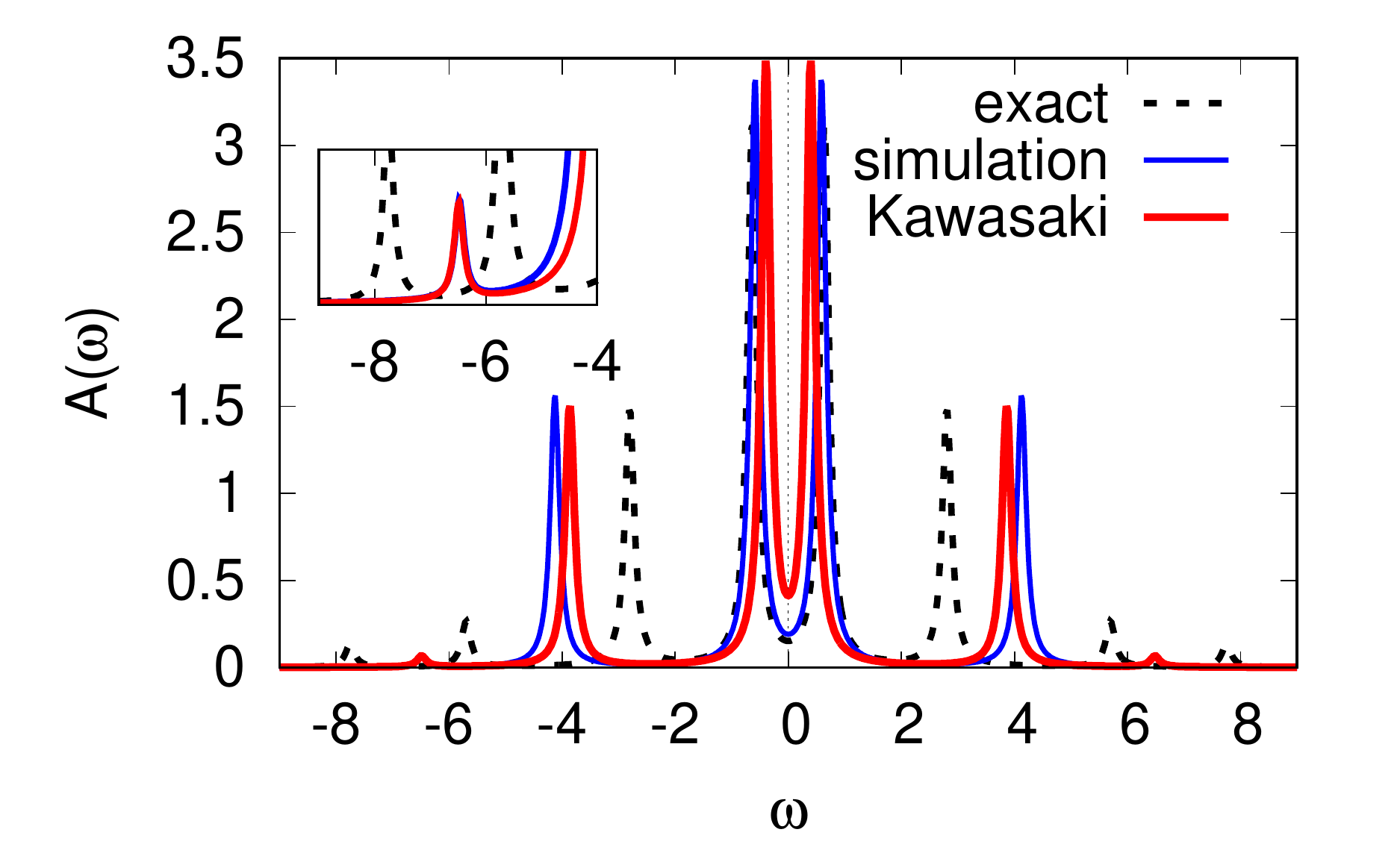}
    \caption{The spectral function of the impurity site obtained on the \textit{Kawasaki} quantum computer (red solid line), compared to the noise-free simulation (blue line) and the exact result with the same boson-cutoff (black dashed line), using a broadening of $\delta=0.1$.  The renormalized bonding-antibonding splitting and the correlation induced satellites at energies around $\pm 3$  are well reproduced. A small plasmon peak is observed at higher energies, with the underestimated weight  being a result of the limitation of the ground state ansatz wave function.
}
    \label{fig:greensfunction}
\end{figure}

In Fig.\ref{fig:greensfunction} we show the resulting impurity Green's function on real frequencies, obtained on the \textit{Kawasaki} quantum computer. The spectrum contains renormalized bonding-antibonding states close to the Fermi level, correlation satellites at intermediate energies and a small plasmonic satellite at higher energies, correctly reproducing the qualitative features of the exact spectral function. Because we calculate the Krylov states up to $n=2$, only first plasmonic satellite can be accessed, while the exact solution has infinitely many satellites of exponentially vanishing weight. The gap around the Fermi level is correctly reproduced on the quantum device, with a splitting of $\Delta=0.9$, underestimating the exact value of $\Delta_{exact}= 1.25$ by a minor degree. The correlation induced satellites at around $\pm 3$ are well reproduced by the KVQA, albeit their energetic position is overestimated ($\pm 3.8$ compared to the exact $\pm 2.8$). Furthermore, on the \textit{Kawasaki} quantum computer we are able to observe the first plasmonic satellite at higher energies in the spectral function at around $\pm 6.5$. Even when the agreement with the exact position and weight is not precise (energetic position overestimated by $1$ and about $25\%$ of the correct weight), this result demonstrates that the KVQA is able to resolve spectral features with relative spectral weight of less than one percent on current NISQ devices.  We want to point out that such a feature would be especially difficult to observe with approaches that obtain the Green's function in real time, like the Trotter expansion or variational methods. The Fourier transform from time to frequency space requires long time scales and fine temporal resolution, which makes it difficult to obtain on current NISQ devices. We thus conclude that the KVQA is especially well suited for obtaining subtle spectral features on noisy quantum devices, as possible errors manifest mostly in the position and weight of the spectral features and thus retain most of the qualitative aspects of the true spectrum.

\begin{figure}[t]
    \centering
    \includegraphics[width=0.5\textwidth]{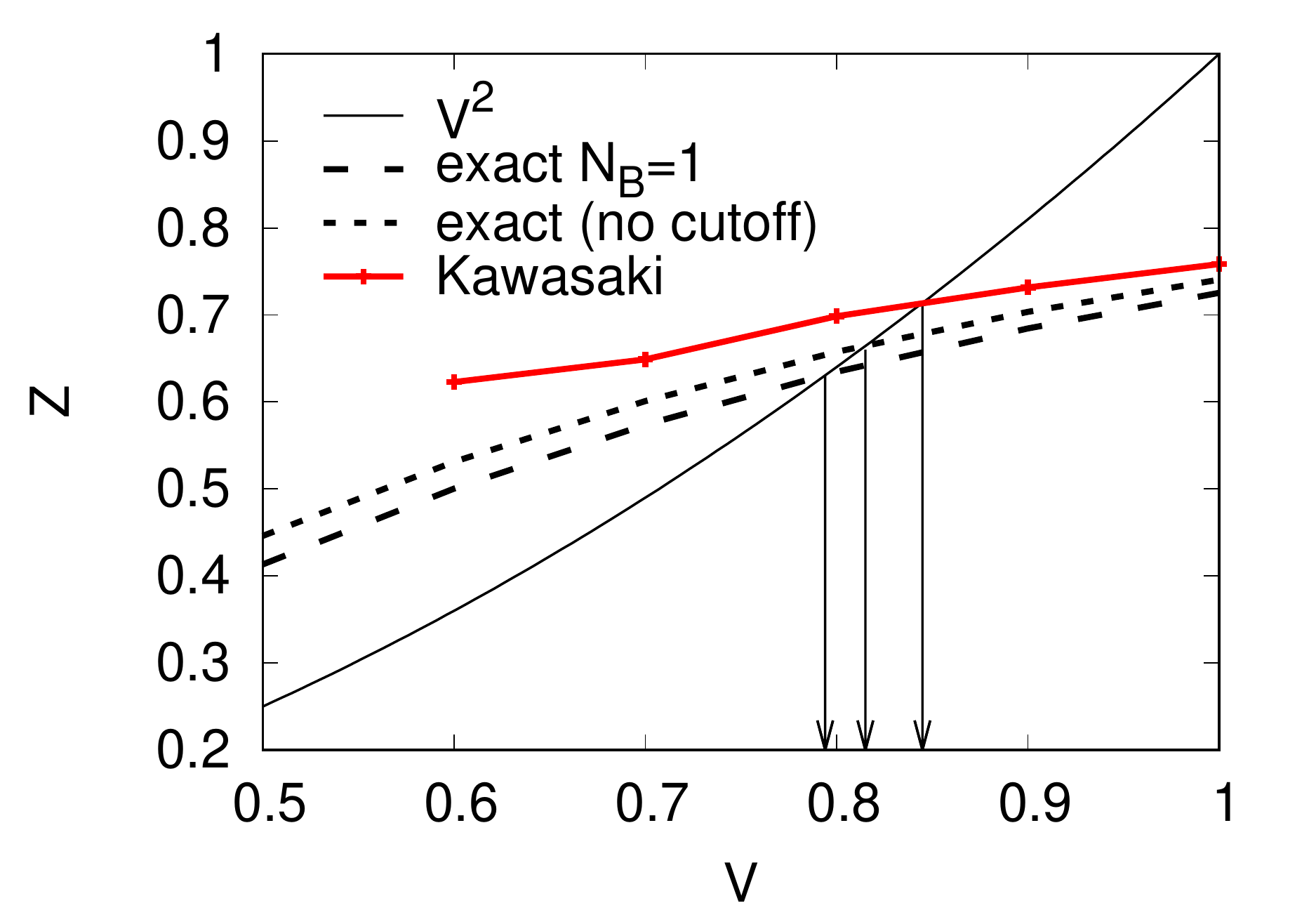}
    \caption{The quasiparticle weight $Z=\left( 1 - \mathrm{Re}\frac{\partial \Sigma}{\partial \omega}\vert_{\omega=0} \right)^{-1}$ obtained from the solution of the Hubbard-Holstein DMFT impurity model for different values of the hybridization strength $V$.  DMFT self-consistency is obtained at $V^2 = Z$, i.e. at the intersection of the two lines indicated by the arrows. The result on the quantum device shows a self-consistency solution around $V=0.84$, close to the exact result with one boson cutoff ($0.79$) and no boson cutoff ($0.81$).
    }
    \label{fig:Zval}
\end{figure}

We now focus on obtaining a solution to the self-consistent DMFT equations for the Hubbard-Holstein model on the Bethe lattice. We set the bandwidth $W=4$, for which the second energy moment of the non-interacting density of states is $M_2=1$. The self-consistency condition is given by $G_{imp}=G_{loc}$, i.e. the impurity Green's function has to be equal to the local lattice Green's function. For the two-site DMFT impurity model at half filling, the bath energy is fixed to the Fermi level, and the self-consistency condition is simplified to a condition for the hybridization amplitude\cite{Potthoff_2siteDMFT}
\begin{align}
    V^2 &= Z M_2,
\end{align}
where $Z=\left( 1 - \mathrm{Re}\frac{\partial \Sigma}{\partial \omega}\vert_{\omega=0} \right)^{-1}$ is the quasiparticle weight, obtained from the derivative of the self-energy $\Sigma = G_0^{-1} - G^{-1}$ on real frequencies. For the parameters considered here, the self-energy shows Fermi liquid behavior\cite{Potthoff_2siteDMFT}, and the quasiparticle weight $Z$ is well defined\footnote{The gap in $A(\omega)$ in Fig.\ref{fig:greensfunction} results from the bonding-antibonding splitting of the two-site model and is not due to a singular self-energy.}. Though, as both the interacting Green's function $G$ and non-interacting Green's function $G_0$ approach zero at $\omega=0$, the derivative of $\Sigma(\omega)$ strongly depends on a precise cancellation of two divergent terms. Similar to previous reports\cite{Keen_dmft_qc} we find this method to be unreliable due to the finite noise level on the quantum device. We therefore obtained $Z$ by integrating the weight of the two peaks closest to the Fermi level. The resulting spectral weight as a function of the hybridization strength $V$ is shown in Fig.\ref{fig:Zval}. The analytical result with a one boson cutoff results in a self-consistent solution at $V=0.79$,  which is close to the exact result  $V=0.81$ without any cutoff on the number of bosons. Comparing to the quasiparticle weight obtained on the \textit{Kawasaki} quantum computer, we observe a good qualitative agreement with the exact solution, albeit the quasiparticle weight is overestimated by about $10$\%. A self-consistent solution can be reliably observed at around $V=0.84$, in reasonable agreement with the exact result.

\begin{figure}[t]
    \centering
    \includegraphics[width=0.5\textwidth]{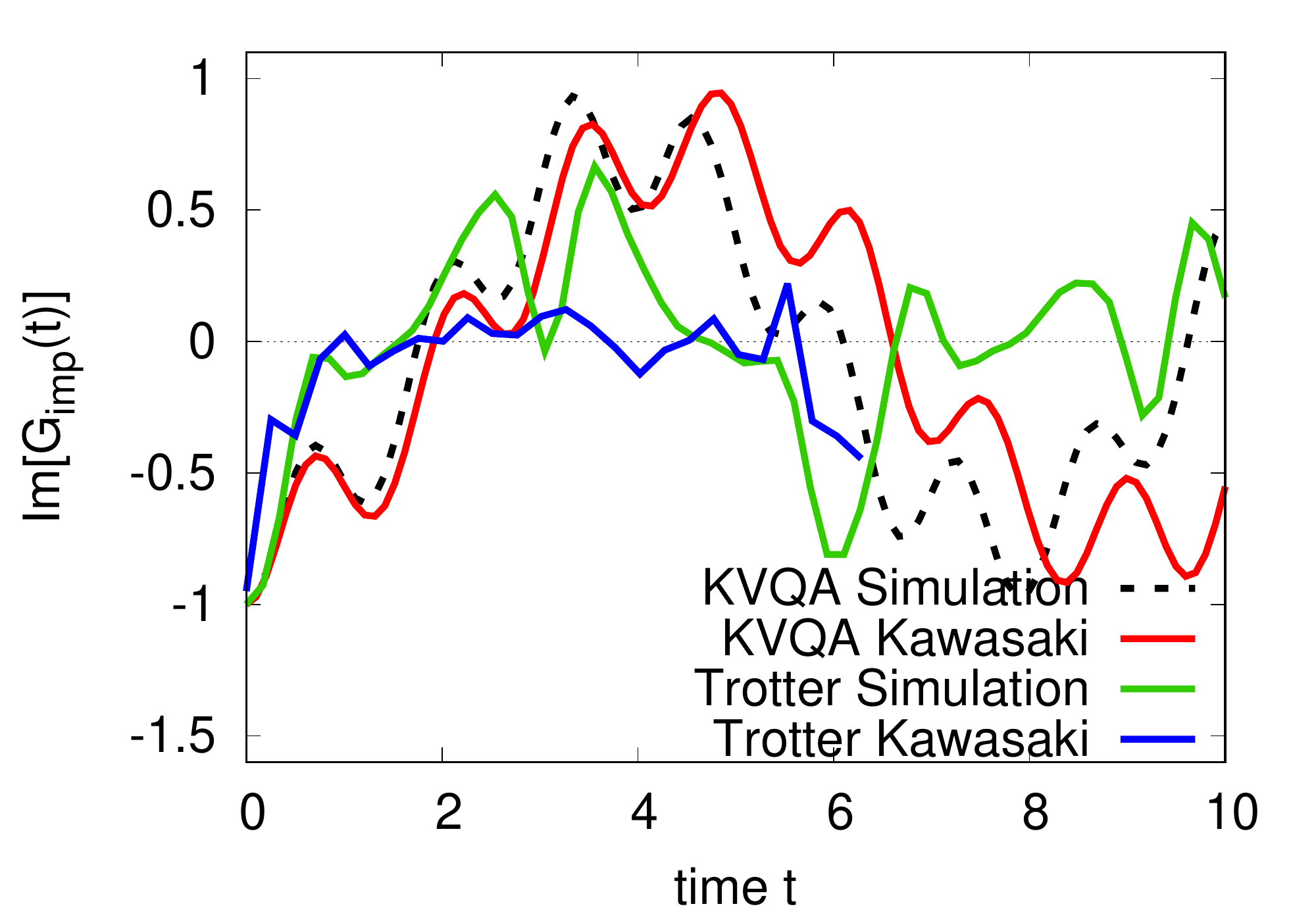}
    \caption{The imaginary part of the impurity Green's function obtained  from a standard Trotter expansion simulation with one Trotter step (green), compared to the Trotter expansion performed on the \textit{Kawasaki} quantum processor (blue solid curve),  and the KVQA  result (red and black curve) for $V=1.0$, $\lambda=1.5$, $\omega_0=5$.     The Trotter expansion provides a much poorer result than the KVQA for two main reasons: The expansion converges slowly with the number of Trotter steps, and it requires much more controlled gate operations than the KVQA, and hence introduces significantly more noise.
    }
    \label{fig:greensfunction_time}
\end{figure}

For comparison we also show the resulting Green's function in the time domain obtained on the \textit{Kawasaki} quantum processor from a Trotter decomposition approach with one Trotter step in Fig.\ref{fig:greensfunction_time}. The Fourier transform of the KVQA Green's function from frequency space to time shows a qualitative agreement with the exact result. In contrast, the Trotter decomposition is significantly less accurate and mostly fluctuates around zero, i.e.  it shows no clear oscillatory behavior and thus provides almost no information about possible spectral features. We identified two main reasons for this: First, the Trotter decomposition converges very slowly for the current system with the number of Trotter steps, therefore, only one step is not sufficient to reproduce the Green's function, as already the simulation without any gate noise is far from the exact result. Second, the Trotter decomposition requires more controlled gate operations than the KVQA which introduce significantly more noise, and thus reduces the quality of the obtained data. Therefore, going beyond one Trotter step and consequently increasing the circuit size would lead to further loss of accuracy. This result shows that the KVQA is significantly more robust and can provide a more accurate result for the Green's function of the DMFT impurity problem for the same quantum computing device.

\section{\label{sec:conclusion}Conclusion}
We have presented an implementation of the Krylov variational quantum algorithm on the IBM \textit{Kawasaki} 27-qubit quantum computer to obtain the Green's function for the Hubbard-Holstein two-site impurity model, and demonstrated that DMFT self-consistency can be reliably obtained. This model extends the Hubbard model and couples the electrons with bosonic degrees of freedom, and is the essential building block for impurity models with frequency dependent interactions such as in extended DMFT, $GW$+DMFT and further nonlocal extensions of DMFT which include nonlocal interactions and dynamical screening effects. We have presented a hardware efficient ansatz for the ground state wave function that exploits the symmetry of the wave function and allows a parametrization in terms of only two parameters. This enabled us to perform a scan of the full two-dimensional parameter space to reliably obtain the ground state and Krylov basis states instead of relying on numerical minimization techniques. The obtained impurity Green's function is in good qualitative agreement with the exact result, exhibiting all major spectral features of the bonding-antibonding, correlation and plasmonic satellites. The approach was shown to be robust enough to reliably obtain the self-consistent solution of the DMFT equations.  We find that the accuracy greatly surpasses previously employed approaches such as the Trotter expansion. Our work forms the basis of future studies of electron-boson coupled systems and nonlocal extensions of DMFT on near-term quantum computers, which are not only important for real materials calculations, but also are computationally very intensive on classical computers, and thus are promising candidates for harnessing the computational capabilities of future quantum computers.

\begin{acknowledgments}
We thank 
E. L\"otstedt,
T. Nishi,
and K. Yamanouchi
for  fruitful discussions and support in utilizing the \textit{Kawasaki} quantum computer at the University of Tokyo. This work is partly supported by the UTokyo Quantum Initiative.
\end{acknowledgments}

%%%%%%%%%%%%%%%%%%%%%%%%%%%%%%%%%%%%%%
\appendix
\section{Implementation details}
\begin{figure}[t]
    \includegraphics[width=0.5\textwidth]{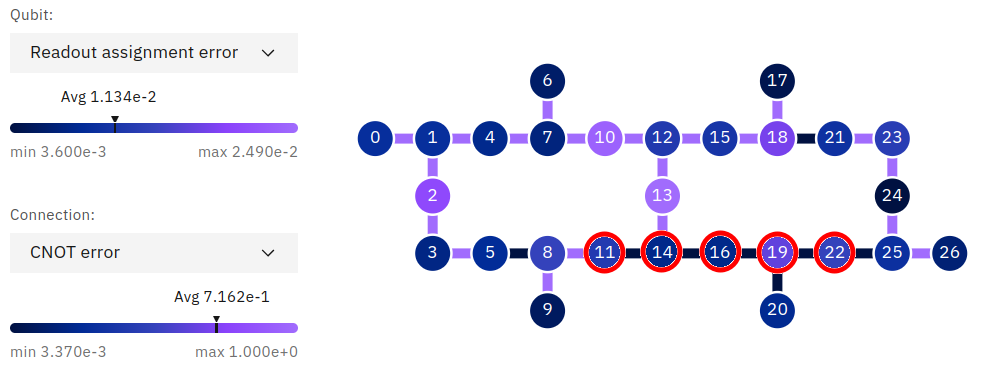}
    \caption{
    The qubits used in this work on the IBM 27-qubit Quantum Falcon Processor \textit{Kawasaki}. 
    The figure is based on the schematic provided by \mbox{https://quantum-computing.ibm.com/services/resources}.   
    }
    \label{fig:qubits}
\end{figure}

For this work we employed the IBM Qiskit environment v. 0.39.2\cite{Qiskit} to implement the quantum circuits and measurements on the IBM 27-qubit Quantum Falcon Processor \textit{Kawasaki}. For all measurements we employed the Sampler primitive, and used the maximum number of shots possible $N=32000$. For the layout we chose the qubits that showed the lowest readout assignment error and were connected with the smallest CNOT error as much as possible, as shown in Fig.\ref{fig:qubits}.

\begin{figure}[t]
    \includegraphics[width=0.23\textwidth]{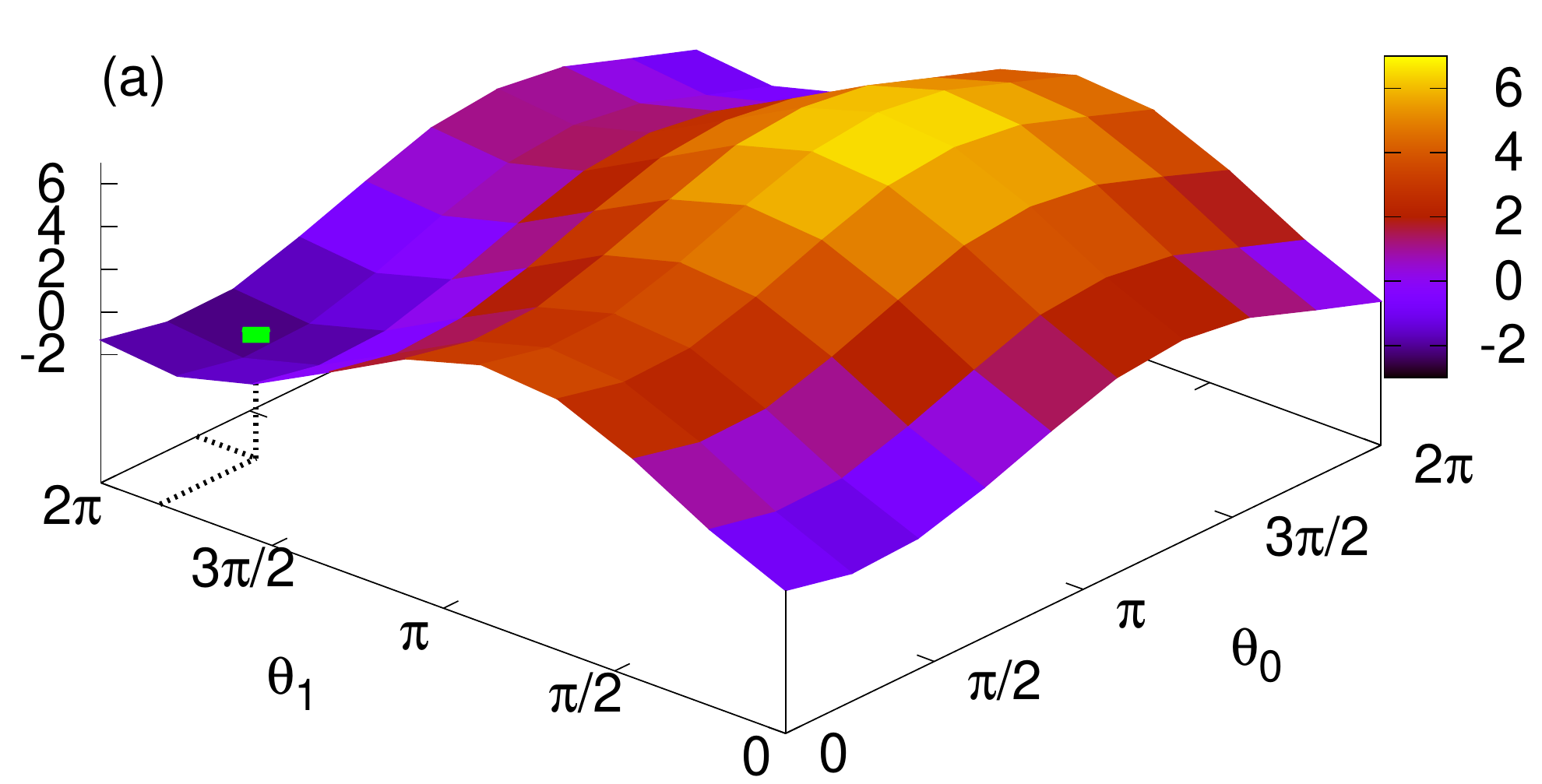}
    \includegraphics[width=0.23\textwidth]{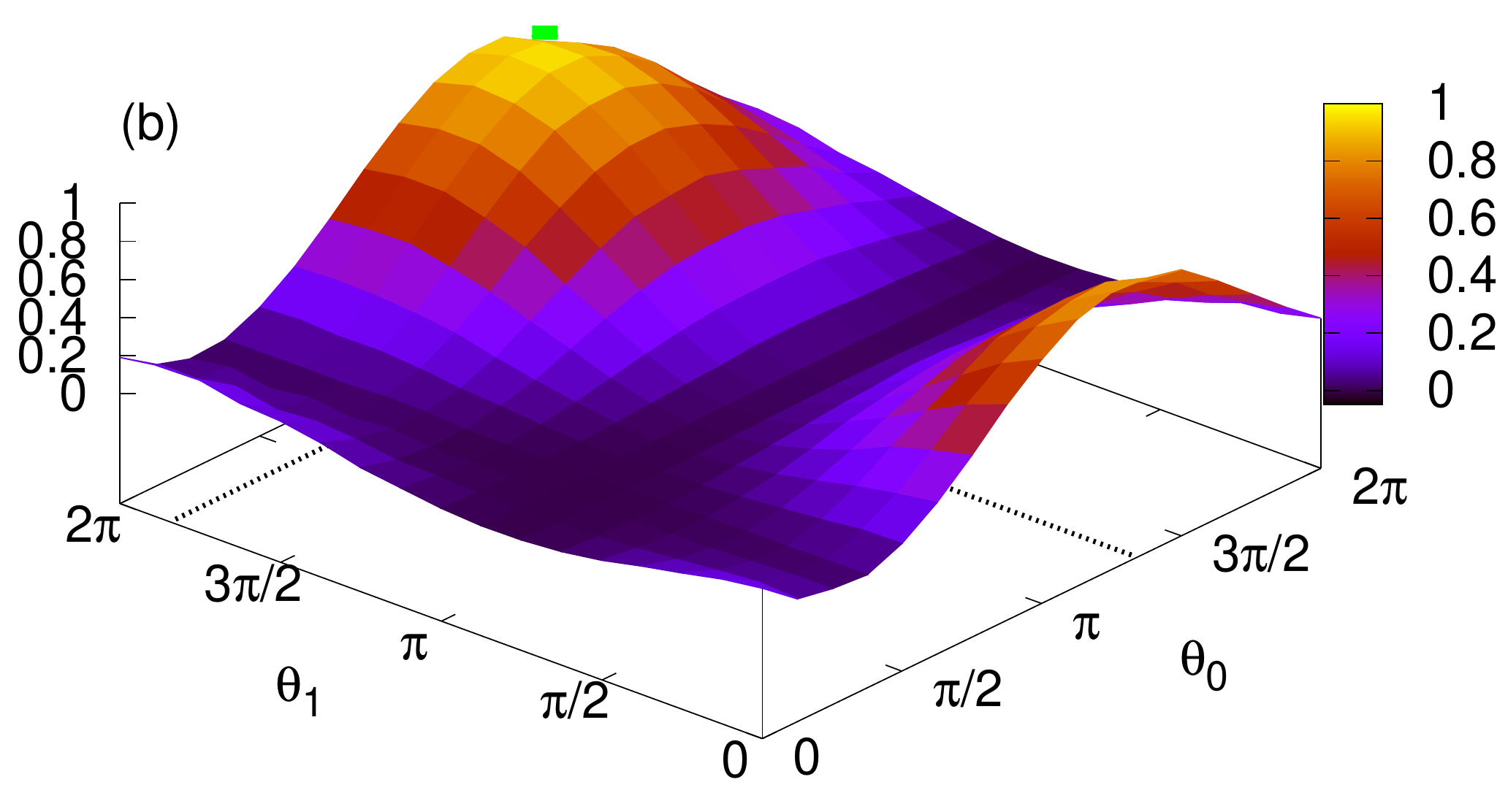}
    \includegraphics[width=0.23\textwidth]{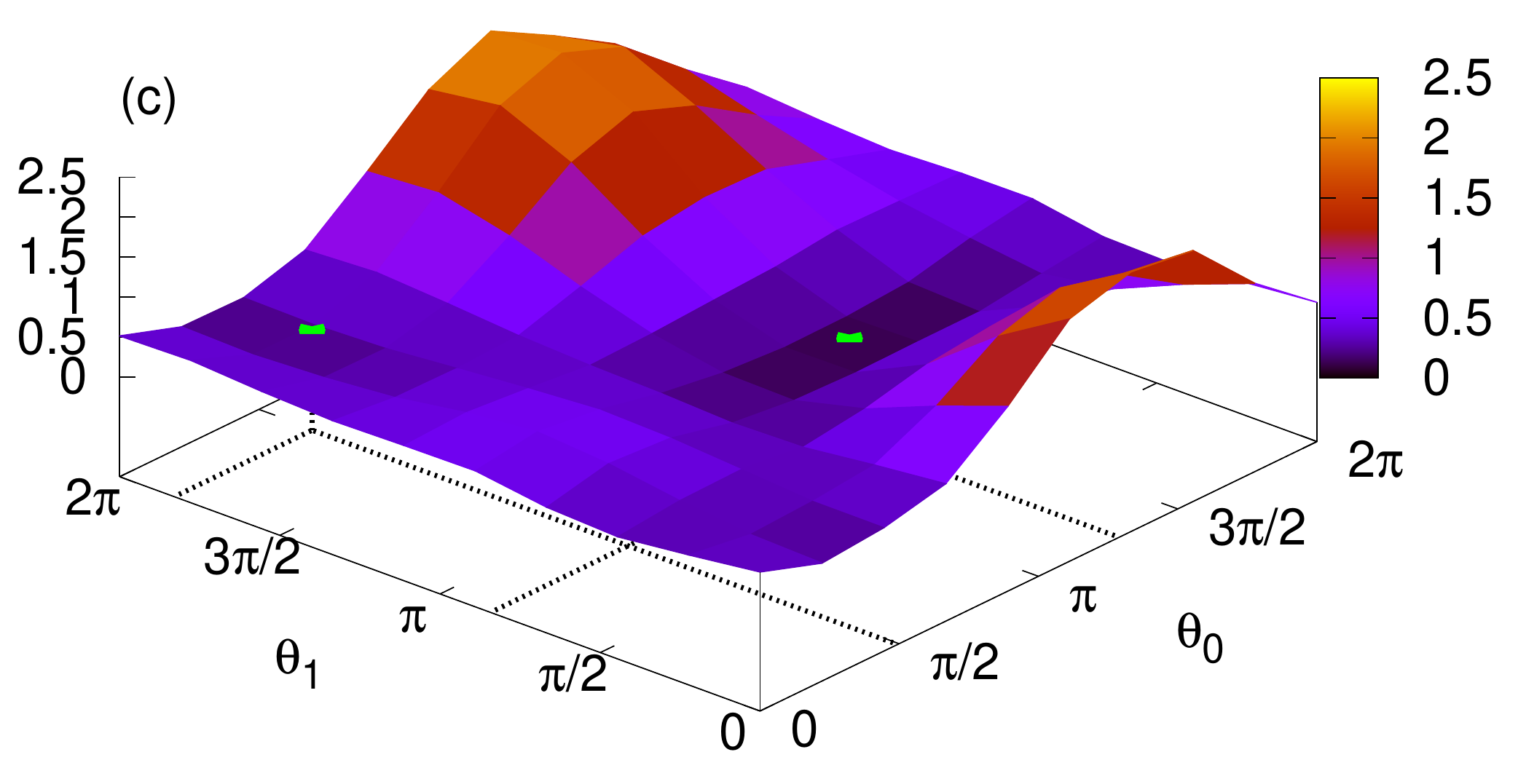}
    \includegraphics[width=0.23\textwidth]{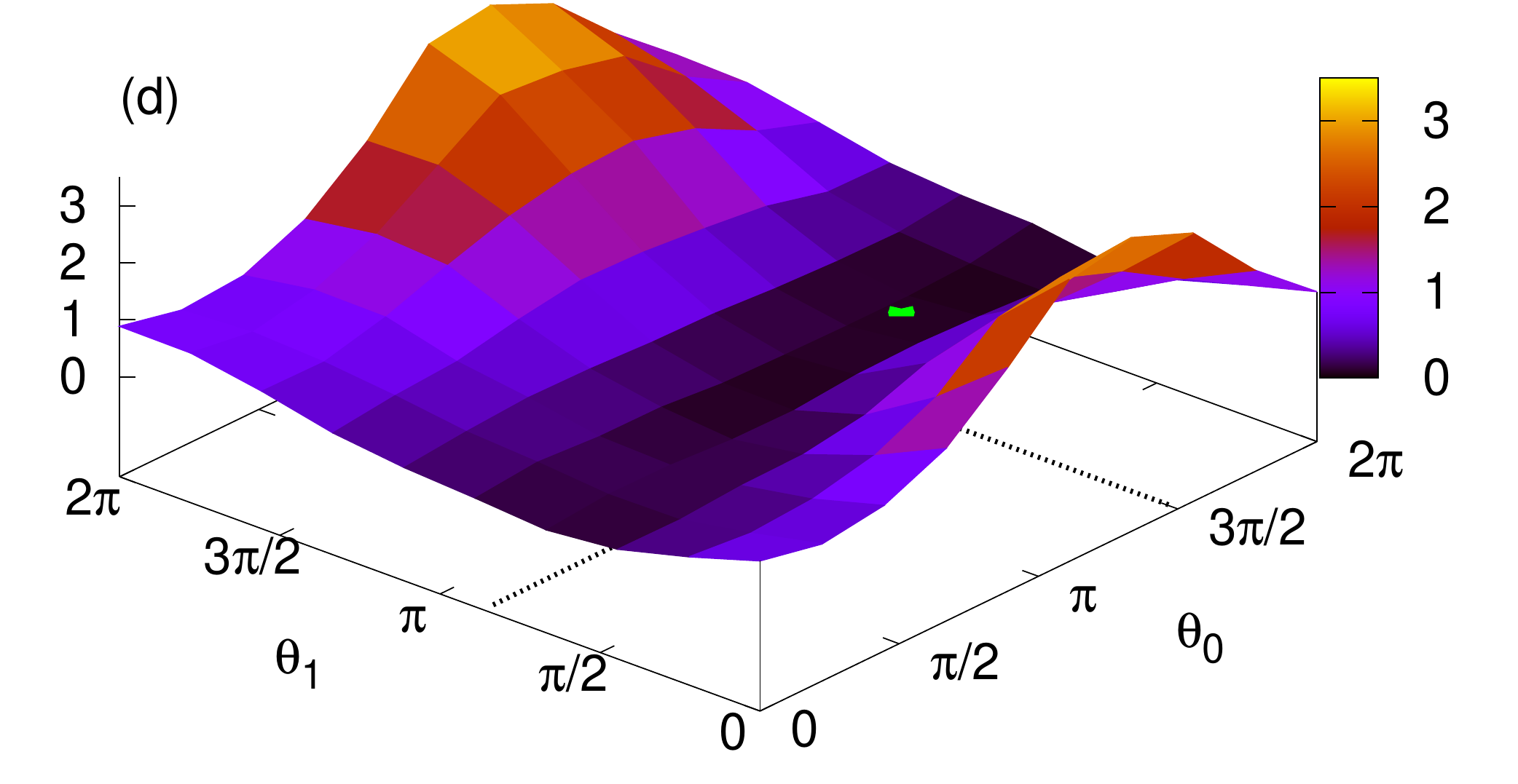}
    \caption{
    The optimization surfaces for the circuit shown in Fig.\ref{fig:vqe_circuit} to find (a) the minimal energy to determine the ground state energy $E_0$ , (b) the maximum overlap to determine the first Krylov vector $| \chi_0 \rangle$, (c) and (d) the minimum of the error function $\sum_i \epsilon_{ni}(\theta)$ to find the next Krylov vectors $| \chi_1 \rangle$  and $| \chi_2 \rangle$, respectively (see Eq.\eqref{eq:eps0}-\eqref{eq:eps2} ).
    }
    \label{fig:optimization}
\end{figure}

To find the Krylov states $| \chi_n \rangle $ in the KVQA we employed the variational method as outlined in Ref.\cite{Jamet2021}: To obtain the parameter set $\{ \theta_n\}$ that generates the state  $| \chi_n \rangle = U(\theta_{n}) | 0 \rangle$ via the set of unitary gate operations $U(\theta_{n}) $  we minimize the three functions
\begin{align}
\epsilon_{n0}( \{ \theta \}) &=  \left(  \frac{| \langle 0|  U^{\dagger}(\theta) H U(\theta_{n-1})  | 0 \rangle  |}{|b_n|} -1\right)^2 , \label{eq:eps0} \\ 
\epsilon_{n1}( \{ \theta \}) &=  | \langle 0|  U^{\dagger}(\theta) U(\theta_{n-1})  | 0 \rangle  |^2 , \\ 
\epsilon_{n2}( \{ \theta \}) &=  | \langle 0|  U^{\dagger}(\theta) U(\theta_{n-2})  | 0 \rangle  |^2 , \label{eq:eps2}
\end{align}
which at the minimum $\epsilon_{ni}=0 \ \forall i$ satisfy  that $\langle \chi_n | H  | \chi_{n-1}  \rangle =b_n$, $\langle \chi_n | \chi_{n-1} \rangle=0$ and $\langle \chi_n | \chi_{n-2} \rangle=0$. 
Specifically we made use of the Trotter-like expansion discussed in the supplementary material I of Ref.\cite{Jamet2021} to evaluate the cost function $\epsilon_{n0}$, which circumvents the introduction of another controlled ancilla qubit. We used 10 steps between $t=0.01...0.3$ to linearly extrapolate the value to $t=0$. In practice we minimize the sum of these three functions, which is the quantity shown in Fig.\ref{fig:optimization} (c) and (d).

Reducing the ansatz wave function to a circuit parametrized by two parameters as shown in Eq.\eqref{eq:ansatz} allowed us to use a two-dimensional scan of the parameter space to determine the approximate ground state and Krylov vectors to obtain the impurity Green's function. In Fig.\ref{fig:optimization} we show the corresponding optimization surfaces obtained on the \textit{Kawasaki} quantum processor. 
The energy surface in Fig.\ref{fig:optimization} (a) showed a clear minimum indicating a unique solution for the approximate ground state. Also the obtained overlap to represent the first Krylov state $|\chi_0\rangle = c | \mathrm{GS}\rangle $ showed a clear maximum with an overlap of more than 0.9 (Fig.\ref{fig:optimization} (b) ).
The optimization surface in Fig.\ref{fig:optimization} (c) for the remaining Krylov states showed two main local minima, corresponding to two possible Krylov vectors. Choosing either local minima leads to the other minima becoming the only global minima in the next iteration, as can be seen in Fig.\ref{fig:optimization} (d). The precise location of the minima was noticeably affected by noise, but we found that the resulting spectral function did not strongly depend on the exact choice of the parameters. While the peak positions and weights of mainly the two outermost peaks (correlation induced and plasmon satellite) at energies above $\pm 3$ were affected to a small degree, the qualitative features of the spectral function stayed the same.

%%%%%%%%%%%%%%%%%%%%%%%%%%%%%%%%%%%%%%
\section{Trotterization and variational approach}
\begin{figure*}[t]
    \includegraphics[width=0.45\textwidth]{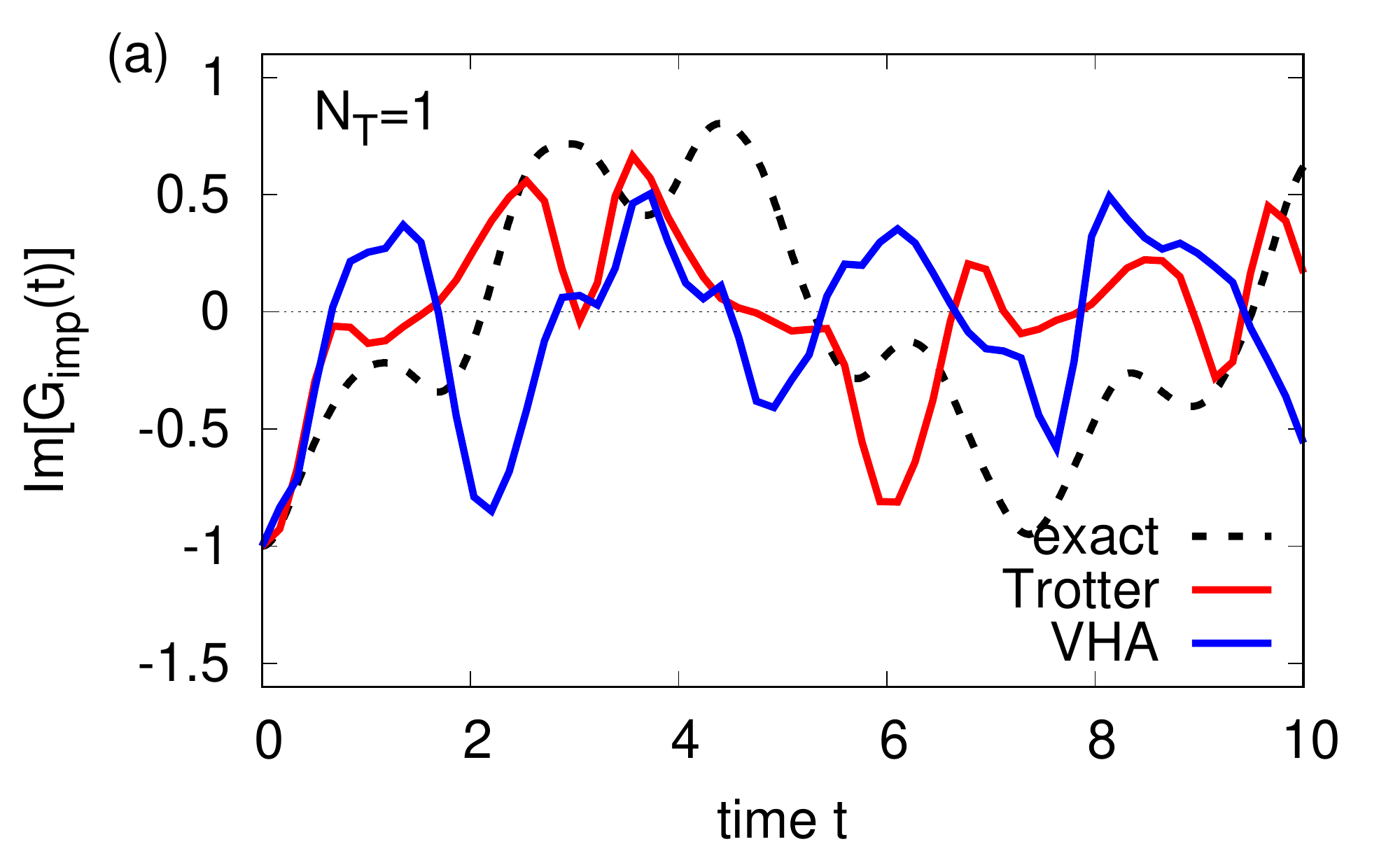}
    \includegraphics[width=0.45\textwidth]{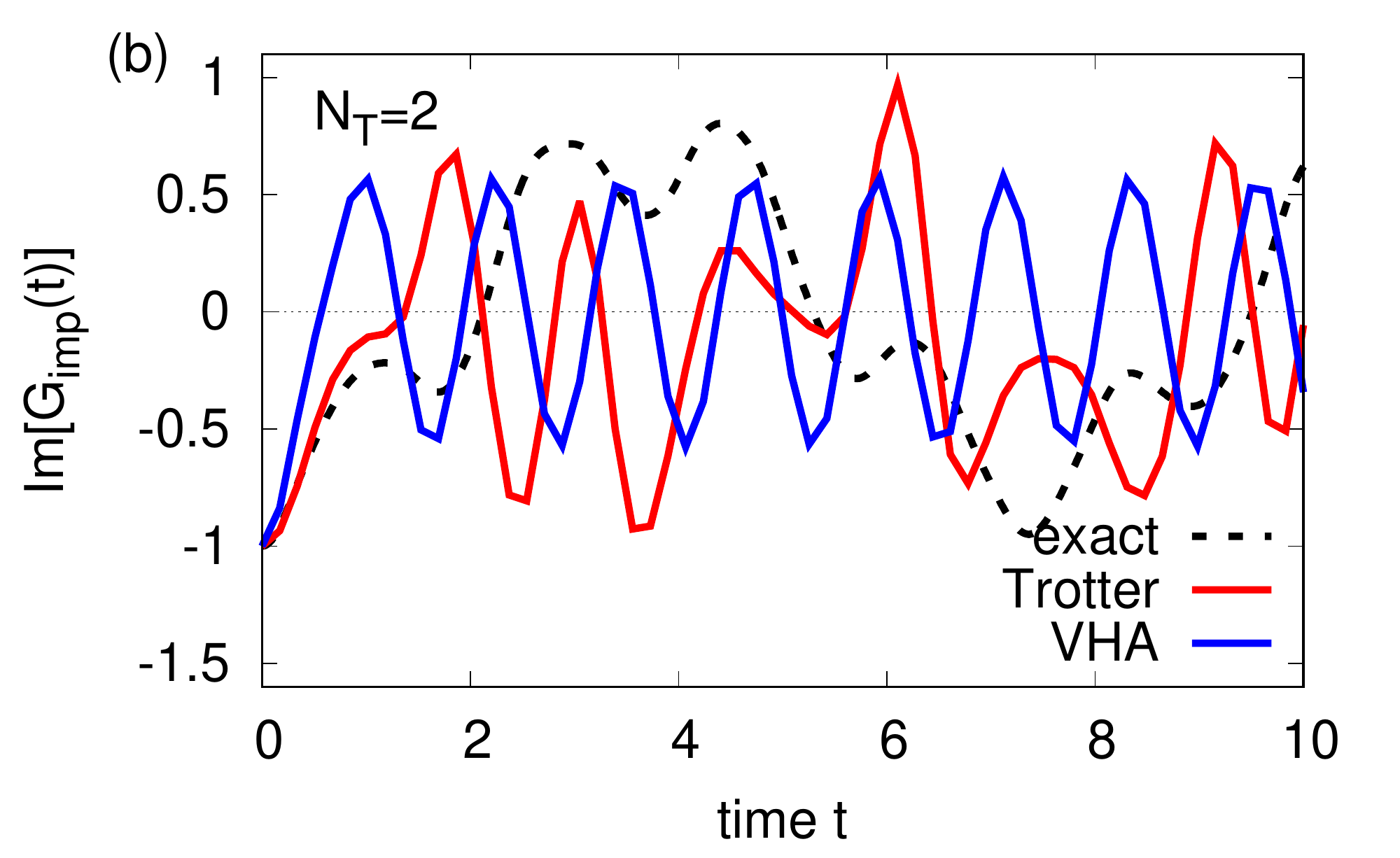}
    \caption{
The imaginary part of the impurity Green's function as a function of time obtained from a Trotter expansion (red line) and variational Hamiltonian ansatz (VHA, blue line) on a classical computer, compared to the exact result for the same boson cutoff $N_B=1$ at $V=1.0$ (other parameters as in the main text). (a)  and (b) show the results for $N_T=1$ and $N_T=2$ Trotterization steps, respectively. Both results are significantly less accurate than the KVQA one shown in Fig.\ref{fig:greensfunction_time}. Increasing the number of Trotter steps to $N_T=2$  results in improvement at small times for the Trotter expansion, but actually worsens the agreement with the exact solution for VHA. 
    }
    \label{fig:trotter_VHA_comparison}
\end{figure*}

As discussed in the main text, we found that the Trotter expansion or the variational Hamiltonian ansatz (VHA)\cite{Wecker2015,Reiner_2019,Libbi2022}, using McLachlan's variational principle as detailed in Ref.~\cite{Libbi2022}, for the two-site DMFT impurity problem show a very poor agreement with the exact result, and converge very slowly with increasing the number of Trotterization steps. To demonstrate this, we show in Fig.\ref{fig:trotter_VHA_comparison} the resulting impurity Green's function obtained from the Trotter expansion and VHA obtained from a classical simulation, for $N_T=1$ and $N_T=2$ Trotterization steps. The agreement with the exact result is very poor, and quickly deviates even at small time scales. Increasing the number of Trotter steps improves the result for the Trotter expansion at small times, but we have found that at least $N_T\sim 8-10$  steps are needed in order to obtain a reasonable agreement up to $T_{max}=10$. Such a large circuit is currently not feasible for present NISQ devices. In contrast to that, the VHA obtains an even worse result when increasing the Trotterization steps to $N_T=2$. We found that the result strongly depends on the operator ordering in the expansion of the time evolution operator
\begin{align}
    U_{VHA} = \prod_{i=1}^{N_T}\left( \prod_{m=1}^{N_H} e^{i\theta_m(t) P_m} \right) ,
\end{align}
where $P_m$ are the Pauli operators of the Hamiltonian after the Jordan-Wigner transformation, with the total number of operator terms $N_H$.
For most orderings the ansatz did not converge to the correct result when increasing the number of Trotter steps.

\bibliography{refs}
\end{document}